%% file: main.tex
\documentclass{webofc}
\usepackage[varg]{txfonts}   
\usepackage{hyperref,amsmath,amssymb,xspace,listings,lstautogobble,tikz,pgfplots}

\definecolor{palette0_0}{HTML}{f6ae2d}
\definecolor{palette0_1}{HTML}{6e1423}
\definecolor{palette0_2}{HTML}{cc444b}
\definecolor{palette0_3}{HTML}{00296b}
\definecolor{palette0_4}{HTML}{00509d}
\pgfplotsset{
    compat=1.3,
    legend style={
        font={\fontsize{2}{0}\selectfont},legend style={row sep=0.2pt},
        anchor=north,at={(0.5,-0.15)},
    },
    legend columns=2,
    every axis/.style={cycle list={{palette0_0,thick,mark=x},
        {palette0_1,thick,mark=triangle},{palette0_2,thick,mark=triangle*},
        {palette0_3,thick,mark=square},{palette0_4,thick,mark=square*},
    }},
    every axis title/.append style={yshift=-1ex},
    grid=both,
    scaled x ticks=false,
    xticklabel style={
        /pgf/number format/fixed
    },
}

\definecolor{blue3}{HTML}{204a87}
\lstdefinestyle{c++}{
    language=c++,
    basicstyle={\ttfamily\footnotesize},
    keywordstyle={\color{blue3}\bfseries},
    commentstyle={\color{black!50}},
    stringstyle={\color{red!50!black}},
}

\usetikzlibrary{fit,matrix,positioning,chains,arrows.meta,shapes.multipart,patterns}
\usetikzlibrary{pgfplots.groupplots}
\pgfdeclarelayer{background}
\pgfsetlayers{background,main}

\title{Exploring Object Stores for High-Energy Physics Data Storage}
\author{\firstname{Javier} \lastname{López-Gómez}\inst{1}\fnsep\thanks{\email{javier.lopez.gomez@cern.ch}} \and
        \firstname{Jakob} \lastname{Blomer}\inst{1}\fnsep\thanks{\email{jblomer@cern.ch}}
}
\institute{CERN, Geneva, Switzerland}
\abstract{%
  Over the last two decades, ROOT TTree has been used for storing over one exabyte of High-Energy Physics (HEP) events. The TTree columnar on-disk layout has been proved to be ideal for analyses of HEP data that typically require access to many events, but only a subset of the information stored for each of them. Future colliders, and particularly HL-LHC, will bring an increase of at least one order of magnitude in the volume of generated data. Therefore, the use of modern storage hardware, such as low-latency high-bandwidth NVMe devices and distributed object stores, becomes more important. However, TTree was not designed to optimally exploit modern hardware and may become a bottleneck for data retrieval. The ROOT RNTuple I/O system aims at overcoming TTree's limitations and at providing improved efficiency for modern storage systems. In this paper, we extend RNTuple with a backend that uses Intel DAOS as the underlying storage, demonstrating that the RNTuple architecture can accommodate high-performance object stores. From the user perspective, data can be accessed with minimal changes to the code, that is by replacing a filesystem path by a DAOS URI. Our performance evaluation shows that the new backend can be used for realistic analyses, while outperforming the compatibility solution provided by the DAOS project.
}
\makeatletter
\hypersetup{pdftitle={\@title},pdfauthor={Javier López-Gómez and Jakob Blomer},allcolors=blue3,colorlinks=true}
\let\@URL=\url\def\url#1{\texttt{\footnotesize\@URL{#1}}} 
\makeatother
\input{defs.tex}

\begin{document}
\maketitle
\input{content/introduction.tex}
\input{content/related_work.tex}
\input{content/background.tex}
\input{content/design.tex}
\input{content/evaluation.tex}
\newpage 
\input{content/conclusion.tex}

\section*{Acknowledgements}
We would like to thank Miguel Fontes Medeiros, Luca Atzori, and Guillermo Izquierdo Moreno from CERN OpenLab for providing access to the DAOS testbed. We would like to thank Johann Lombardi, Mohamad Chaarawi, and Andrea Luiselli from Intel for the DAOS support and recommendations regarding the design and optimization.

\bibliography{main}

\end{document}

%% file: defs.tex
\newcommand{\RNTuple}{\textsf{RNTuple}\xspace}
\newcommand{\TTree}{\textsf{TTree}\xspace}
\newcommand{\RPageSource}{\texttt{RPageSource}\xspace}
\newcommand{\RPageSink}{\texttt{RPageSink}\xspace}

\newcommand{\dfuse}{\textsf{dfuse}\xspace}
\newcommand{\dkey}{\textit{dkey}\xspace}
\newcommand{\akey}{\textit{akey}\xspace}

%% file: content/introduction.tex
\section{Introduction}\label{sec:introduction}
In High-Energy Physics (HEP), an event is encoded as a record that may contain a number of variable-length collections or properties. For instance, an event might contain a collection of particles, each including a set of scalar properties (e.g., momentum, energy, etc.), a collection of tracks, jets, and any other data collected by a particle detector. Most analysis of HEP data require access to many events, but only require a subset of the properties stored for each instance. In this scenario, using conventional row-oriented storage systems is suboptimal, as they incur in an overhead due to reading a high volume of unneeded data.

ROOT \TTree\cite{brun1997root} avoids this overhead by using a columnar layout, i.e., consecutively storing values of the same property for a range of rows. This encoding not only avoids the aforementioned overhead, but also contributes to improve data compression as similar values are stored together. While \TTree has been used to efficiently store more than 1\,EB of HEP data during the last two decades, it was not designed to fully exploit modern hardware and/or storage systems, e.g. NVMe devices and object stores. In a previous publication~\cite{blomer2020evolution} we presented \RNTuple, the ROOT's new, experimental columnar I/O subsystem. \RNTuple is a backwards-incompatible redesign of \TTree that overcomes its limitations and takes advantage of state-of-the-art hardware and storage systems. The layered design of \RNTuple permits the separation of encoding and storage of pages (groups of values belonging to the same data column). This separation is crucial for supporting different storage systems, such as POSIX files or object stores.

In this paper, we extend \RNTuple with a new backend that leverages Intel DAOS~\cite{DAOS} as the underlying storage. Given the importance of object stores in next-generation HPC centers, e.g. the Argonne's Aurora~\cite{stevens2019aurora} supercomputer, we aim to bridge the gap between HEP and the upcoming supercomputing facilities. In particular, we contribute with the following:
\begin{itemize}
    \item We provide the design and prototype implementation of a \RNTuple backend that can leverage an existing DAOS deployment to store HEP data. In that regard, we propose and evaluate two different page to object mappings.
    \item We evaluate the read/write performance of our backend with a realistic LHCb analysis in several conditions, and compare it to the DAOS \dfuse compatibility layer and a local filesystem.
\end{itemize}

The rest of this paper is organized as follows. Section~\ref{sec:related_work} briefly reviews other related work found in the literature. Section~\ref{sec:background} provides an introduction to \RNTuple and Intel DAOS. Section~\ref{sec:design} describes the design of the new \RNTuple DAOS backend. In Section~\ref{sec:evaluation}, we evaluate its performance compared to the DAOS \dfuse compatibility layer and a local filesystem. Section~\ref{sec:conclusion} summarizes the contribution and enumerates future work.

%% file: content/related_work.tex
\section{Related work}\label{sec:related_work}
Due to the aforementioned advantages, columnar storage systems are popular not only in high-energy physics but also in other areas. During the last few years, many solutions for columnar storage of large datasets such as Amazon Redshift~\cite{redshift}, Apache Parquet~\cite{Vohra2016} and Apache Arrow~\cite{ApacheArrow} have emerged. Oftentimes, users need to store nested complex data types, which is especially relevant for the HEP area. A well-known portable, cross-platform solution for this is HDF5~\cite{10.1145/1966895.1966900}; however, it currently does not support columnar storage.

At the time of this writing, we found few publications in the literature targeting DAOS integration for next-generation exascale HPC centers. In particular, HDF5 already implements a connector for DAOS~\cite{HDF5_DAOS}. Also, there is work in progress to integrate DAOS with Apache Parquet/Arrow, and the Lustre~\cite{braam2019lustre} parallel filesystem.

%% file: content/background.tex
\section{Background}\label{sec:background}
In this section, we briefly describe the ROOT project and its experimental, new columnar I/O subsystem (\RNTuple). Additionally, we introduce Intel DAOS (Distributed Asynchronous Object Store)~\cite{DAOS}, a contemporary object store relevant for next-generation HPC centers.

\subsection{The ROOT project and RNTuple}\label{sec:root_rntuple}
ROOT~\cite{brun1997root} is an open-source data analysis framework that is part of the software stack of many HEP experiments, e.g.~CERN's ATLAS and CMS. ROOT provides components for statistics analysis, data visualization, convenient data storage/retrieval, and is bundled with an interactive C++ interpreter that can be used for fast prototyping of analysis code. ROOT also includes dynamic Python bindings that provide access to C++ types and objects. The traditional columnar HEP data storage is provided by \TTree.

\RNTuple is the new columnar I/O subsystem that shall replace \TTree in the long term. \RNTuple has been designed to take advantage of state-of-the-art hardware and storage systems. Its implementation makes thorough use of template metaprogramming and features available in recent C++ standards, which contributes to a lower run-time overhead and a more maintainable and simpler code. Several design decisions were taken to improve the effective read/write bandwidth and decrease the size of the data stored on a persistent medium. For instance, integer and floating-point numbers use little-endian encoding so that they match the endianness of most popular architectures, which allows for direct memory-mapping of pages; booleans are packed into octets, and more compression schemes are under investigation. \RNTuple was previously described in \cite{blomer2020evolution}. For the sake of space, we will only summarize the high-level architecture.

\begin{figure}[htb]
    \centering
    \resizebox{0.64\textwidth}{!}{\input{fig/rntuple-layers.tex}}
    \caption{Layers of the \RNTuple subsystem.}\label{fig:rntuple_layers}
\end{figure}
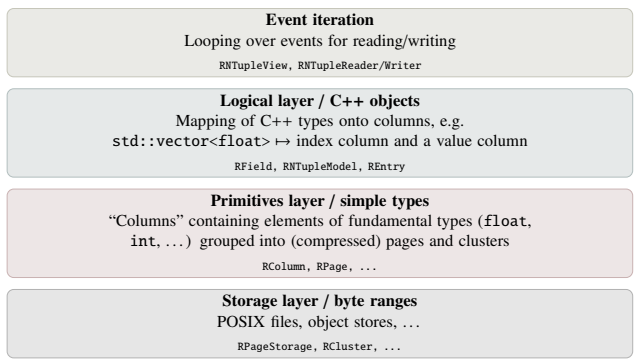

The design of \RNTuple comprises four layers: storage layer, primitives layer, logical layer, and event iteration layer (see Figure~\ref{fig:rntuple_layers}). The storage layer abstracts the details of the underlying storage class (e.g. a POSIX filesystem, an object store, etc.), optionally providing additional packing and/or compression; therefore, this layer is responsible for storing/retrieving byte ranges organized in clusters and pages. Pages contain a range of values for a given column, whereas a cluster contains pages that store data for a specific row range. The primitives layer groups a range of elements of a fundamental type, i.e.~a horizontal partition, into pages and clusters. The logical layer maps complex C++ types to columns of fundamental types, e.g. \verb|std::vector<float>| is mapped to an index column and a value column. Finally, the event iteration layer provides a convenient interface for iterating over events.

\begin{figure}[htb]
    \centering
    \resizebox{0.96\textwidth}{!}{\input{fig/rntuple-format.tex}}
    \caption{\RNTuple on-disk physical layout. A page stores a range of values for the corresponding struct member (same color). Dotted arrows symbolize a reference to a different page.}\label{fig:rntuple_format}
\end{figure}
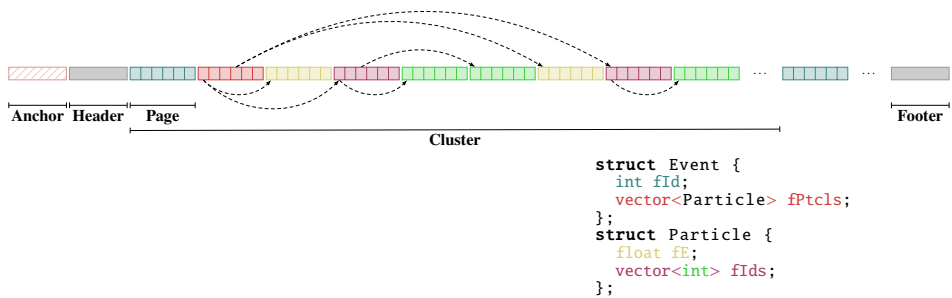

Given this layer organization, support for a new storage system can be added by extending the storage layer via a backend that is able to handle read/write operations for pages and clusters. \RNTuple already includes a file backend that allows reading and writing ROOT files. The relevant on-disk layout is depicted in Figure~\ref{fig:rntuple_format}. This layout comprises an anchor (that includes the offsets and size of the header and footer sections, among other information), a header (that describes the schema), a footer (that encodes locations of pages/clusters), and a number of pages grouped into clusters.

\subsection{Intel DAOS}\label{sec:daos}
Traditional storage systems and kernel I/O schedulers were designed for high-latency rotating disks that could handle a few hundred IOPS (I/O operations per second). On the one hand, optimizations applied in the block layer (coalescing, buffering, etc.) are less relevant for modern storage devices, such as NVMe SSDs and Intel Optane DC persistent memory, and they add overhead if used. On the other hand, the POSIX I/O semantics and particularly its strong consistency model, have been identified as a major problem in parallel filesystem scalability. To tackle this issue, Intel DAOS provides a fault-tolerant object store optimized for high bandwidth, low latency, and high IOPS, that targets next-generation exascale HPC centers~\cite{stevens2019aurora} and aims at fully exploiting modern hardware. Traditionally, object stores provide at least the \textit{get} and \textit{put} primitives (\textit{fetch} and \textit{update} in DAOS, respectively) to access object content.

\begin{figure}[htb]
    \centering
    \input{fig/daos-poolcontobj.tex}
    \caption{Graphical representation of different DAOS entities.}\label{fig:daos_contpoolobj}
\end{figure}
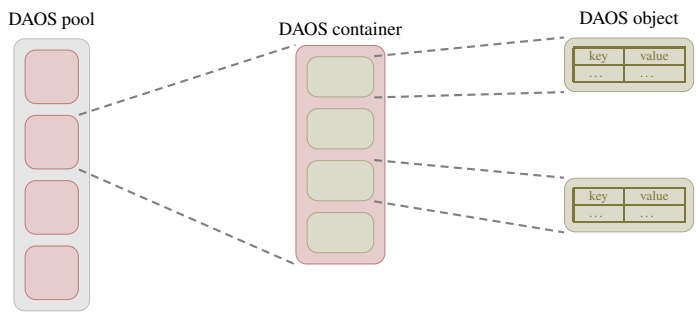

In DAOS, a system is comprised of a number of servers that run a Linux daemon exporting local NVM storage. DAOS servers listen on one management interface and many fabric endpoints (for data transport). Storage resources are partitioned into \textit{targets} that can be accessed independently to avoid contention. The system administrator can create pools that may span a number of servers. Its storage is distributed among the available targets; such a division is called a \textit{shard}. Containers can be created by users using the space assigned to a pool. Objects can then be placed into a container. Pools and containers are identified by a UUID (Universally Unique IDentifier), while objects can be referenced using a 128-bit object identifier (OID). An object is a key-value store with locality and redundancy/replication (as determined by the object class). The key is split into distribution key (\dkey) and attribute key (\akey), where values for the same \dkey will be co-located in the same target. Object contents can be accessed through three different APIs: multi-level key-array (native), key-value, and one-dimensional array. The relationship between the different entities can be seen in Figure~\ref{fig:daos_contpoolobj}.

DAOS also provides components that permit to port applications. Specifically, it implements a POSIX filesystem that can be used either through \texttt{libioil} (I/O interception library) or \texttt{dfuse} (a FUSE filesystem), a ROMIO driver for applications using MPI-IO, and connectors for other systems such as HDF5 and Apache Spark.

%% file: fig/rntuple-layers.tex
\tikz[start chain=c going above,node distance=1.5ex,
      layer/.style={on chain,rounded corners,draw=#1!30!black!30,fill=#1!30!black!10,text width=\columnwidth,align=center}
]{
  \node[layer=black]{{\bf Storage layer / byte ranges}\\
    POSIX files, object stores, \ldots\\[0.5ex]
    {\tt\scriptsize RPageStorage, RCluster, \ldots}};

  \node[layer=red]{{\bf Primitives layer / simple types}\\
    ``Columns'' containing elements of fundamental types (\texttt{float}, \texttt{int}, \ldots) grouped into (compressed) pages and clusters\\[0.5ex]
    {\tt\scriptsize RColumn, RPage, \ldots}};

  \node[layer=teal]{{\bf Logical layer / C++ objects}\\
    Mapping of C++ types onto columns, e.g. \texttt{std::vector\textless float\textgreater} $\mapsto$ index column and a value column\\[0.5ex]
    {\tt\scriptsize RField, RNTupleModel, REntry}};

  \node[layer=olive]{{\bf Event iteration}\\
    Looping over events for reading/writing\\[0.5ex]
    {\tt\scriptsize RNTupleView, RNTupleReader/Writer}};
}

%% file: fig/rntuple-format.tex
\tikz[start chain=c going right,node distance=2pt,
      section/.style={minimum width=46pt,minimum height=10pt,inner sep=2pt},
      page/.style={section,rectangle split,rectangle split parts=6,rectangle split horizontal,
        draw=#1!76!black!76,fill=#1!76!black!20},
      header/.style={section,draw=black!40,fill=black!20},
      footer/.style={section,draw=black!40,fill=black!20},
      ntuple anchor/.style={section,draw=red!67!black!40,pattern=north east lines,pattern color=red!67!black!20},
      >={Latex[length=1ex]},
]{
  \path node(anchor) [on chain,ntuple anchor]{}
    node(header) [on chain,header]{};
  \foreach \i [count=\c] in {teal,red,yellow,purple,green,green,yellow,purple,green}
    \node(page-\c) [on chain,page=\i]{};
  \path node(ldots-1) [on chain,minimum width=30pt]{\ldots}
    node [on chain,page=teal]{}
    node [on chain,minimum width=30pt]{\ldots};
  \node(footer) [on chain,footer]{};

  \def\pointer#1#2#3{
    \draw[->,thick,densely dashed,bend right=#3] (#1) to (#2);
  }
  \foreach \i/\j in {page-2/page-3,%
        page-2/page-4,%
        page-4/page-5,%
        page-8/page-9}
    \pointer{\i.one south}{\j.one south}{45};
  \pointer{page-2.four north}{page-7.one north}{-30};
  \pointer{page-2.four north}{page-8.one north}{-30};
  \pointer{page-4.three north}{page-6.one north}{-30};

  \def\annotate#1#2#3#4{
    \draw[|-|] ([yshift=#1]#2.south west) -- ([yshift=#1]#3.south east) node [midway,below,font={\bfseries\fontsize{13}{13}\selectfont}] {#4};
  }
  \annotate{-20pt}{anchor}{anchor}{Anchor}
  \annotate{-20pt}{header}{header}{Header}
  \annotate{-20pt}{page-1}{page-1}{Page}
  \annotate{-40pt}{page-1}{ldots-1}{Cluster}
  \annotate{-20pt}{footer}{footer}{Footer}

  \node(lst) [below=0.74in of page-6,xshift=2in] {%
  \begin{lstlisting}[style=c++,basicstyle={\ttfamily\fontsize{14}{14}\selectfont},keywordstyle={\bfseries}]
  struct Event {
    `\color{teal!76!black!76}int fId`;
    `\color{red!76!black!76}vector$<$`Particle`\color{red!76!black!76}$>$ fPtcls`;
  };
  struct Particle {
    `\color{yellow!76!black!76}float fE`;
    `\color{purple!76!black!76}vector$<$\color{green!76!black!76}int\color{purple!76!black!76}$>$ fIds`;
  };
\end{lstlisting}};
}

%% file: fig/daos-poolcontobj.tex
\tikz[container/.style 2 args={rounded corners,draw=#1!black!50,fill=#1!black!20,inner sep=4pt,label={above:#2}},
    every label/.style={font=\scriptsize},
    element/.style n args={3}{rounded corners,draw=#1!black!50,fill=#1!black!20,minimum width=#2,minimum height=#3,font=\tiny}
]{
    \def\zoomin#1#2{\draw[densely dashed,thick,black!50] (#1.north east) -- (#2.north west)
        (#1.south east) -- (#2.south west);}

    \begin{scope}[start chain=c going below,node distance=4pt]
        \foreach \i in {1,...,4} \node[on chain,element={red!50}{2em}{2em}] {};
        \begin{pgfonlayer}{background}
            \node(P) [container={white!50}{DAOS pool},fit={(c-1)(c-4)}] {};
        \end{pgfonlayer}
    \end{scope}

    \begin{scope}[start chain=o going below,node distance=4pt,xshift=1.5in]
        \foreach \i in {1,...,4} \node[on chain,element={olive!50}{2.5em}{1.5em}] {};
        \begin{pgfonlayer}{background}
            \node(C) [container={red!50}{DAOS container},fit={(o-1)(o-4)}] {};
        \end{pgfonlayer}
    \end{scope}
    \zoomin{c-2}{C}

    \begin{scope}[start chain=O going below,node distance=32pt,xshift=3in,yshift=1ex]
        \foreach \i in {1,2} \node[on chain,element={olive!50}{4em}{2em},inner ysep=1pt] {\color{olive!70!black!96}\begin{tabular}{|l|l|}\hline
        key     & value     \\\hline
        \ldots  & \ldots    \\\hline\end{tabular}};
        \node[above=0pt of O-1,font=\scriptsize] {DAOS object};
    \end{scope}
    \zoomin{o-1}{O-1}\zoomin{o-3}{O-2}
}

%% file: content/design.tex
\section{Design of the RNTuple DAOS backend}\label{sec:design}
This section explains in detail the new \RNTuple DAOS backend. In Section~\ref{sec:overview}, we give an overview of the design and engineering decisions. In Section~\ref{sec:mapping}, we describe how clusters and pages map to DAOS objects. Finally, Section~\ref{sec:user_interface} illustrates required changes to the user code.

\subsection{Overview}\label{sec:overview}
\RNTuple relies on the \RPageSource and \RPageSink classes for reading and writing pages, respectively. Therefore, new backends can be implemented by inheriting from these classes. The \RPageSource class defines two member functions that can be overridden in a subclass: \texttt{PopulatePage()}, responsible for reading a single page, and \texttt{LoadCluster()}, that is called to read all pages belonging to a given cluster. The latter can optionally implement additional optimizations, e.g. vector reads and parallel decompression. The \RPageSink class provides the \texttt{CommitPage()} and \texttt{CommitCluster()} methods that are responsible for writing data. Like their read counterparts, they may include optimizations to coalesce write requests.

The C DAOS API is provided by libdaos. While it can be directly consumed in a C++ program, it involves writing boilerplate code and requires manual resource management. To hide away complexity and manage shared resources, we wrapped a subset of libdaos functionality into three C++ classes: \texttt{RDaosPool}, \texttt{RDaosContainer}, and \texttt{RDaosObject}. \texttt{RDaosPool} handles the connection to a DAOS pool. This connection is kept open at least until all the containers have been closed. \texttt{RDaosContainer} gives basic access to objects inside a container; additionally, it provides vector read/write of multiple objects via \texttt{ReadV()} and \texttt{WriteV()}. Internally, these functions rely on libdaos event queues and asynchronous fetch/update operations. Finally, low-level access to objects is provided by the \texttt{RDaosObject} class.

In summary, \texttt{PopulatePage()} and \texttt{CommitPage()} execute a synchronous read or write operation through the simple interface defined by \texttt{RDaosContainer}. Additionally, reading all the pages in a cluster is optimized using a vector read. At the time of this writing, vector writes are not used. The scheme for assigning OIDs, \dkey, and \akey is described in the next section.

Finally, we provide mocks for the libdaos subset used by the previous classes that makes it possible to test the backend in those environments lacking a real DAOS deployment.

\subsection{Mapping RNTuple clusters and pages to objects}\label{sec:mapping}
Given the richness of the DAOS interface, several mappings of RNTuple pages and clusters to DAOS entities are conceivable.

Our proposal defines two possible mappings for pages (and clusters, where applicable) which are described in the following.
\begin{description}
    \item[One OID per page.] A sequential OID is assigned for each committed page. We make no specific use of \dkey and \akey, i.e., both are kept constant. This na\"ive association does not fully exploit DAOS capabilities.

    \item[One OID per cluster.] The OID equals the \RNTuple clusterId; the \dkey is a sequential number that is used for addressing pages in the cluster; \akey has no specific use and is kept constant. This impacts how pages are distributed among targets in DAOS servers, which might improve both read and write bandwidth.
\end{description}

Regardless of the mapping, independent OIDs and constant \dkey/\akey pair are used to store the \RNTuple header, footer, and anchor. We compared the read/write throughput measured for both cases as part of our experimental evaluation in Section~\ref{sec:evaluation}.

\subsection{User interface}\label{sec:user_interface}
The \texttt{RNTupleReader} and \texttt{RNTupleWriter} classes provide a convenient interface to store/retrieve data. The second and third arguments of the constructor specify, respectively, the n-tuple name and the location. This location is typically a filesystem path; however, a DAOS container may be referenced through the use of a specific URI format. These URIs contain, in order, the string `\texttt{daos://}', the pool UUID followed by `\texttt{:}', the list of service replica ranks separated by `\texttt{\_}', and the container UUID. As can be seen in Figure~\ref{fig:rntuplereader_usage}, user code only requires minor changes to read or write data in a DAOS container.

The previous schema allows a user to access a DAOS container transparently. Nevertheless, it requires the manual specification of the pool and container UUIDs. While this is not user-friendly, we consider finding the physical path of a dataset a task of the data management system and outside the scope of RNTuple.

\begin{figure}[htb]
    \centering
    \input{fig/rntuplereader-usage.tex}
    \caption{Changes required to the user code. Note the use of a \texttt{daos://} URI in sub-figure (b).}\label{fig:rntuplereader_usage}
\end{figure}

%% file: fig/rntuplereader-usage.tex
\begin{minipage}[b]{0.48\textwidth}
    \centering
\begin{lstlisting}[style=c++,basicstyle={\ttfamily\tiny}]
auto ntuple = RNTupleReader::Open(
    "DecayTree",
    "./B2HHH~zstd.ntuple");

auto viewH1IsMuon = ntuple->GetView<int>("H1_isMuon");
`\ldots`
\end{lstlisting}
    \mbox{\footnotesize(a) File-backed storage.}
\end{minipage}%
\begin{minipage}[b]{0.48\textwidth}
    \centering
\begin{lstlisting}[style=c++,basicstyle={\ttfamily\tiny}]
auto ntuple = RNTupleReader::Open(
    "DecayTree",
    "`\color{red}daos://4b614f30-f476-4831-84ba-a51197600020:1/f1b0a25a-7fbb-`
          `\color{red}4fba-b7d2-9a9f4e10e8f4`");

auto viewH1IsMuon = ntuple->GetView<int>("H1_isMuon");
`\ldots`
\end{lstlisting}
    \mbox{\footnotesize(b) DAOS-backed storage.}
\end{minipage}

%% file: content/evaluation.tex
\section{Evaluation}\label{sec:evaluation}
In this section, we present the results of the experimental evaluation of the \RNTuple DAOS backend we carried out. Specifically, we measured performance differences w.r.t.~file-backed storage in a variety of situations. All the experiments in this performance evaluation employed the following hardware and software environment.

\begin{description}
    \def\indent{\rule{1em}{0pt}}
    \item[Hardware platform.] We used the CERN Openlab DAOS testbed to run the experiments. This cluster comprises three servers and two client nodes, connected to an Omni-Path Edge 100 Series 24-port switch. Note that we only made use of one client node, i.e., all tests ran on the same host. The hardware specifications for the cluster are detailed in the following paragraphs.

    \indent\textit{Server nodes.} These nodes have $2\times$ Intel Xeon Platinum 8260 CPU (24 physical cores) running at 2.4\,GHz, 35.75\,MB of L3 cache, 384\,GB of DDR4 RAM. Note that HyperThreading was enabled (total of 96 logical cores per node). The nodes are equipped with $12\times$ 128\,GB DDR4 Optane DCPMM memory, $4\times$ 1\,TB NVMe SSDs (Intel SSDPE2KX010T8), and an Intel Omni-Path HFI Silicon 100 Series Host Fabric Adapter. The OS is CentOS Linux 7.9 (kernel 3.10.0-1127).

    \indent\textit{Client node.} The client node is equipped with $2\times$ Intel Xeon Platinum 8160 CPU (24 physical cores) running at 2.1\,GHz, and 192\,GB of DDR4 RAM; HyperThreading is enabled for this node (total of 96 logical cores). High-bandwidth interconnection is provided by a Intel Omni-Path HFI Silicon 100 Series Host Fabric Adapter. The OS is CentOS Linux 7.9 (kernel 3.10.0-1160).

    \item[Software.] The evaluation was carried out using DAOS 0.9.4, libfabric 1.7.2, libpsm2 11.2.78, and ROOT revision \texttt{1f1e9b8}. DAOS was configured to use PSM2 transport (\texttt{ofi+psm2}) and flow control was disabled (\texttt{CRT\_CREDIT\_EP\_CTX=0}) in order to avoid limiting the bandwidth of in-flight asynchronous requests.

    \item[Test cases.] In order to evaluate the performance of our proposal, we used the \texttt{gen\_lhcb} and \texttt{lhcb} programs from the RNTuple benchmark repository~\cite{iotools}. These programs are used to import an existing CERN LHCb dataset (in a ROOT file) into \RNTuple, and to perform a typical LHCb analysis, respectively. In the following, we will use \texttt{gen\_lhcb}  to measure the write throughput. Similarly, \texttt{lhcb} provides a measurement of the read throughput.\par
    We evaluated the performance of the \RNTuple libdaos-based backend w.r.t.~the use of a local filesystem and the \dfuse FUSE filesystem. In the subsequent plots, the data point and error bars correspond, respectively, to the average and minimum/maximum value over 20 executions of the same test. Specifically, we analyzed the performance in the following scenarios.
    
    \indent\textit{Constant page size, increasing cluster size.} The number of elements per page is kept constant at $10\,000$, while the number of elements per cluster is doubled on each iteration (starting at $20\,000$). A page size of $10\,000$ roughly equates the \TTree default basket size of 32\,kB.
    Because \RNTuple loads in parallel pages that belong to the same cluster, this scenario evaluates the effect of queuing many small read operations.

    \indent\textit{Increasing page size, constant cluster size.} The number of elements per page is doubled on each iteration (starting at $10\,000$), while the number of elements per cluster is kept constant at $320\,000$.
    In this case, we aim at measuring the impact of the I/O request size on the throughput.
\end{description}

In the following, we provide an analysis of the read/write performance for the previous test cases in a variety of configurations. Specifically, we measured differences in using different DAOS object classes (SX and RP\_XSF) and compression algorithms (none and zstd). The SX object class forces spreading across all the targets in a pool. Alternatively, RP\_XSF uses many replicas so that it is highly scalable for fetching data. The local filesystem used for the tests is XFS on an Intel Optane NVMe SSD. 

\subsection{Performance analysis for one OID per page}
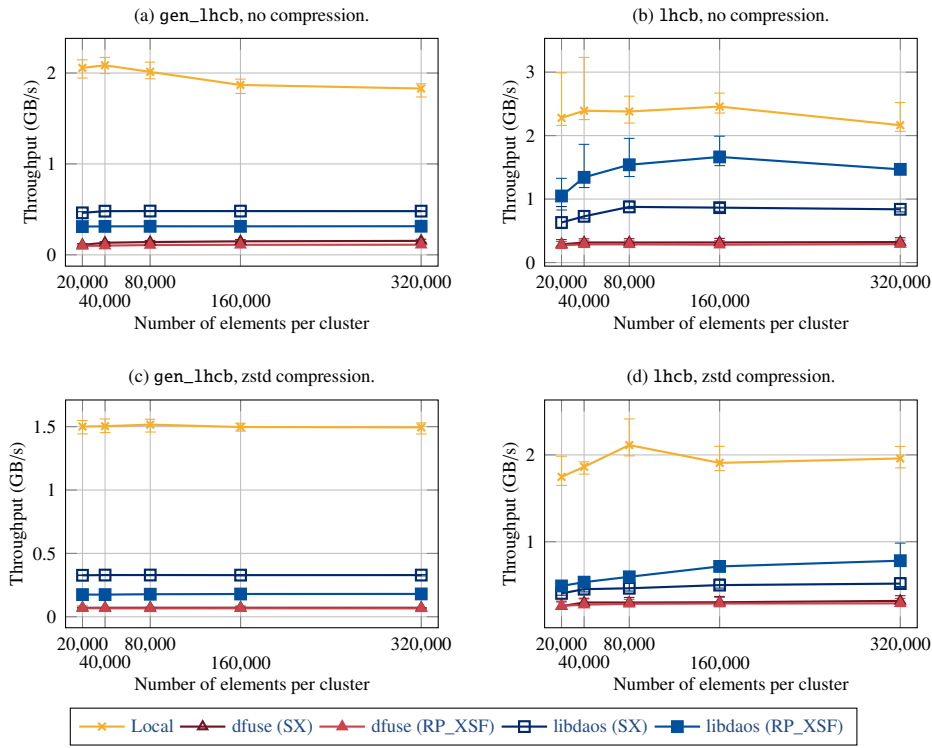
\begin{figure}[htb]
    \centering
    \input{fig/oidperpage_cPSiCS.tex}
    \vspace{-1em}
    \caption{One OID per page, constant page size and increasing cluster size. The \texttt{lhcb} and \texttt{gen\_lhcb} programs measure the read or write throughput, respectively.}\label{fig:oidperpage_cPSiCS}
\end{figure}

In this section, we evaluate the read/write performance for the one OID per page case. The size of the LHCb dataset for these tests is 1.5\,GB. Figure~\ref{fig:oidperpage_cPSiCS} shows the throughput obtained for different cluster sizes while the page size is kept constant, i.e., the number of pages per cluster that can be read in parallel is increasingly larger. As can be seen in Fig.~\ref{fig:oidperpage_cPSiCS}.b and Fig.~\ref{fig:oidperpage_cPSiCS}.d, the libdaos-based backend peaks at $1.66$\,GB/s for $160\,000$ elements per cluster. Given a small page size of less than 80\,kB, a larger cluster size is detrimental for reading, i.e., issuing many small I/O requests negatively impacts the read throughput. Also, using the RP\_XSF object class results in noticeable benefits, as it increases the read throughput at the cost of a slightly degraded write performance (see Fig.~\ref{fig:oidperpage_cPSiCS}.a and Fig.~\ref{fig:oidperpage_cPSiCS}.b).

\begin{figure}[htb]
    \centering
    \input{fig/oidperpage_iPScCS.tex}
    \vspace{-1em}
    \caption{One OID per page, increasing page size and constant cluster size. The \texttt{lhcb} and \texttt{gen\_lhcb} programs measure the read or write throughput, respectively.}\label{fig:oidperpage_iPScCS}
\end{figure}
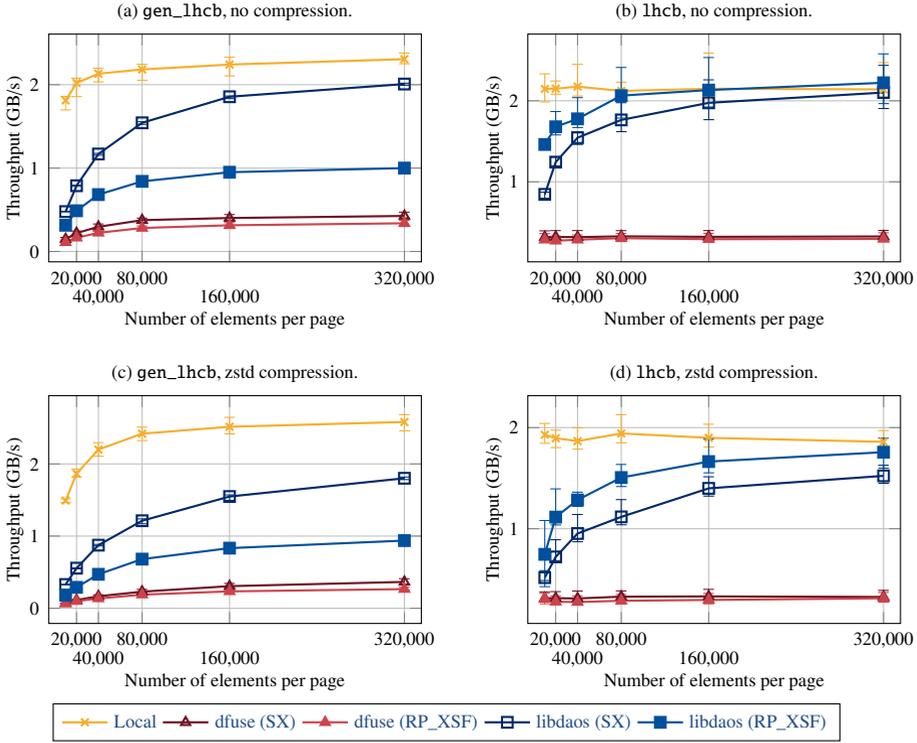

Figure~\ref{fig:oidperpage_iPScCS} shows the measured throughput in case of increasing the page size while keeping the cluster size constant at $320\,000$. As can be observed, the DAOS backend benefits from using larger I/O sizes which contributes to improve the read/write throughput or even surpass local SSDs. It is also worth mentioning that the proposed backend outperformed \dfuse in all situations. Despite the promising results (best case is 2.22\,GB/s), the read throughput is still far below the 12.3\,GB/s measured by the IOR~\cite{shan2007using} benchmark tool using a transfer size of 4\,MB and 16 processes.

\subsection{Performance analysis for one OID per cluster}
\begin{figure}[htb]
    \centering
    \input{fig/oidpercluster_iPScCS.tex}
    \vspace{-1em}
    \caption{Comparison of OID/page and OID/cluster, increasing page size and constant cluster size ($320\,000$). The \texttt{lhcb} and \texttt{gen\_lhcb} programs measure the read or write throughput, respectively.}\label{fig:oidpercluster_iPScCS}
\end{figure}
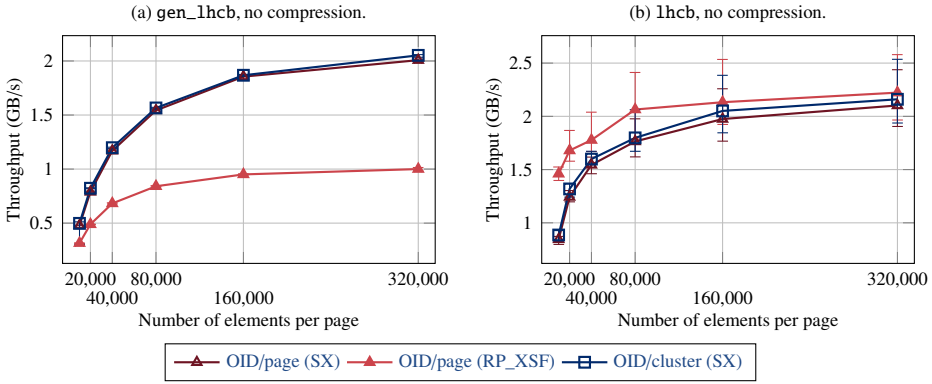

In this section, we compare the performance for one OID per cluster w.r.t.~one OID per page. Figure~\ref{fig:oidperpage_iPScCS} shows the impact of larger reads on the performance. As of $160\,000$ elements per page, i.e. around 1\,MB reads, the remote I/O to DAOS matches the speed of a local NVMe device. As can be seen in Figure~\ref{fig:oidpercluster_iPScCS}, using one OID per cluster slightly improves the read throughput for a given object class. An explanation for this is that having a non-constant distribution key leads to a more even distribution of data among targets. However, the benefits seen in the write throughput are negligible in this case. Note that measurements for ``OID/cluster (RP\_XSF)'' were prohibited by a known problem in the libpsm2 library.

%% file: fig/oidperpage_cPSiCS.tex
\tikz[every node/.style={font=\scriptsize}]{
    \begin{groupplot}[group style={group size=2 by 2,horizontal sep=40pt,vertical sep=50pt},
    width=0.5\textwidth,height=1.8in]
        \gdef\datA{data/genlhcb_cPSiCS_none.dat}\gdef\lblA{(a) \texttt{gen\_lhcb}, no compression.}
        \gdef\datB{data/lhcb_cPSiCS_none.dat}   \gdef\lblB{(b) \texttt{lhcb}, no compression.}
        \gdef\datC{data/genlhcb_cPSiCS_zstd.dat}\gdef\lblC{(c) \texttt{gen\_lhcb}, zstd compression.}
        \gdef\datD{data/lhcb_cPSiCS_zstd.dat}   \gdef\lblD{(d) \texttt{lhcb}, zstd compression.}
        
        \pgfplotsinvokeforeach {A,B,C,D} {
            \nextgroupplot[title={\csname lbl#1\endcsname},enlarge x limits=0.05,legend style={legend columns=-1,legend to name=gplegend_5_1},
            xlabel={Number of elements per cluster},ylabel={Throughput (GB/s)},
            every axis x label/.append style={yshift=1ex},every axis y label/.append style={yshift=-1ex},
            xtick={20000,40000,80000,160000,320000},
            error bars/y dir=both,error bars/y explicit,
            x tick label style={yshift={-mod(\ticknum,2)*0.7em}}
            ]
            \addplot table [x=CS,y=localfs,y error minus=localfs_m,y error plus=localfs_M] {\csname dat#1\endcsname};                     \addlegendentry{Local}
            \addplot table [x=CS,y=dfuse_SX,y error minus=dfuse_SX_m,y error plus=dfuse_SX_M] {\csname dat#1\endcsname};                  \addlegendentry{dfuse (SX)}
            \addplot table [x=CS,y=dfuse_RP_XSF,y error minus=dfuse_RP_XSF_m,y error plus=dfuse_RP_XSF_M] {\csname dat#1\endcsname};      \addlegendentry{dfuse (RP\_XSF)}
            \addplot table [x=CS,y=libdaos_SX,y error minus=libdaos_SX_m,y error plus=libdaos_SX_M] {\csname dat#1\endcsname};            \addlegendentry{libdaos (SX)}
            \addplot table [x=CS,y=libdaos_RP_XSF,y error minus=libdaos_RP_XSF_m,y error plus=libdaos_RP_XSF_M] {\csname dat#1\endcsname};\addlegendentry{libdaos (RP\_XSF)}
        }
    \end{groupplot}
    \node at ($(group c1r2) + (1.1in,-1.1in)$) {\ref{gplegend_5_1}};
}

%% file: fig/oidperpage_iPScCS.tex
\tikz[every node/.style={font=\scriptsize}]{
    \begin{groupplot}[group style={group size=2 by 2,horizontal sep=40pt,vertical sep=50pt},
    width=0.5\textwidth,height=1.8in]
        \gdef\datA{data/genlhcb_iPScCS_none.dat}\gdef\lblA{(a) \texttt{gen\_lhcb}, no compression.}
        \gdef\datB{data/lhcb_iPScCS_none.dat}   \gdef\lblB{(b) \texttt{lhcb}, no compression.}
        \gdef\datC{data/genlhcb_iPScCS_zstd.dat}\gdef\lblC{(c) \texttt{gen\_lhcb}, zstd compression.}
        \gdef\datD{data/lhcb_iPScCS_zstd.dat}   \gdef\lblD{(d) \texttt{lhcb}, zstd compression.}
        
        \pgfplotsinvokeforeach {A,B,C,D} {
            \nextgroupplot[title={\csname lbl#1\endcsname},enlarge x limits=0.05,legend style={legend columns=-1,legend to name=gplegend_5_1},
            xlabel={Number of elements per page},ylabel={Throughput (GB/s)},
            every axis x label/.append style={yshift=1ex},every axis y label/.append style={yshift=-1ex},
            xtick={20000,40000,80000,160000,320000},
            error bars/y dir=both,error bars/y explicit,
            x tick label style={yshift={-mod(\ticknum,2)*0.7em}}
            ]
            \addplot table [x=PS,y=localfs,y error minus=localfs_m,y error plus=localfs_M] {\csname dat#1\endcsname};                     \addlegendentry{Local}
            \addplot table [x=PS,y=dfuse_SX,y error minus=dfuse_SX_m,y error plus=dfuse_SX_M] {\csname dat#1\endcsname};                  \addlegendentry{dfuse (SX)}
            \addplot table [x=PS,y=dfuse_RP_XSF,y error minus=dfuse_RP_XSF_m,y error plus=dfuse_RP_XSF_M] {\csname dat#1\endcsname};      \addlegendentry{dfuse (RP\_XSF)}
            \addplot table [x=PS,y=libdaos_SX,y error minus=libdaos_SX_m,y error plus=libdaos_SX_M] {\csname dat#1\endcsname};            \addlegendentry{libdaos (SX)}
            \addplot table [x=PS,y=libdaos_RP_XSF,y error minus=libdaos_RP_XSF_m,y error plus=libdaos_RP_XSF_M] {\csname dat#1\endcsname};\addlegendentry{libdaos (RP\_XSF)}
        }
    \end{groupplot}
    \node at ($(group c1r2) + (1.1in,-1.1in)$) {\ref{gplegend_5_1}};
}

%% file: fig/oidpercluster_iPScCS.tex
\tikz[every node/.style={font=\scriptsize}]{
    \begin{groupplot}[group style={group size=2 by 2,horizontal sep=40pt,vertical sep=50pt},
    width=0.5\textwidth,height=1.8in]
        \gdef\datA{data/genlhcb_iPScCS_none_OIDcluster.dat}\gdef\lblA{(a) \texttt{gen\_lhcb}, no compression.}
        \gdef\datB{data/lhcb_iPScCS_none_OIDcluster.dat}   \gdef\lblB{(b) \texttt{lhcb}, no compression.}

        \pgfplotsinvokeforeach {A,B} {
            \nextgroupplot[title={\csname lbl#1\endcsname},enlarge x limits=0.05,legend style={legend columns=-1,legend to name=gplegend_5_2},
            xlabel={Number of elements per page},ylabel={Throughput (GB/s)},
            every axis x label/.append style={yshift=1ex},every axis y label/.append style={yshift=-1ex},
            xtick={20000,40000,80000,160000,320000},
            error bars/y dir=both,error bars/y explicit,
            x tick label style={yshift={-mod(\ticknum,2)*0.7em}},
            cycle list={
                {palette0_1,thick,mark=triangle},{palette0_2,thick,mark=triangle*},
                {palette0_3,thick,mark=square}
            }]
            \addplot table [x=PS,y=OIDpage_SX,y error minus=OIDpage_SX_m,y error plus=OIDpage_SX_M] {\csname dat#1\endcsname};            \addlegendentry{OID/page (SX)}
            \addplot table [x=PS,y=OIDpage_RP_XSF,y error minus=OIDpage_RP_XSF_m,y error plus=OIDpage_RP_XSF_M] {\csname dat#1\endcsname};\addlegendentry{OID/page (RP\_XSF)}
            \addplot table [x=PS,y=OIDcluster_SX,y error minus=OIDcluster_SX_m,y error plus=OIDcluster_SX_M] {\csname dat#1\endcsname};   \addlegendentry{OID/cluster (SX)}
        }
    \end{groupplot}
    \node at ($(group c1r1) + (1.1in,-1.1in)$) {\ref{gplegend_5_2}};
}

%% file: content/conclusion.tex
\section{Conclusion}\label{sec:conclusion}
In this paper, we present an extension to \RNTuple which allows users to store HEP data in a DAOS container, therefore closing the gap between the HEP community and the next-generation HPC centers. From the user point of view, switching to a different storage backend can be performed with minimal efforts, only requiring to change a filesystem path to an URI. As demonstrated in the evaluation, applications benefit of using the dedicated backend, which outperforms the DAOS \dfuse filesystem in all cases. Nevertheless, we were unable to reach the expected peak performance as reported by the IOR benchmark.

As a future work, we aim at investigating the performance issues found in the experimental evaluation. Also, we plan to analyze the impact of an alternative mapping that makes use of the attribute key, i.e., the OID and distribution key shall be derived from the \RNTuple cluster index and column identifier, respectively, while the attribute key is used to address individual pages. Finally, we plan to provide an efficient mechanism to transfer data from HEP storage to DAOS.